\documentstyle[prd,aps,epsf]{revtex}
\draft

\newcommand{\bee}{\begin{equation}}
\newcommand{\ee}{\end{equation}}
\newcommand{\beea}{\begin{eqnarray}}
\newcommand{\eea}{\end{eqnarray}}

\begin{document}
\title{
Simple Observables from Fat Link Fermion Actions
}
\author{T. DeGrand (the MILC collaboration)}
\address{
Physics Department, 
        University of Colorado, 
        Boulder, CO 80309 USA \\
}
\date{\today}

\maketitle

\abstract
A comparison  is made of the (quenched) light hadron
 spectrum and of simple matrix elements
for a hypercubic fermion action (based on a fixed point action)
 and the clover action, both using fat links,
 at a lattice spacing $a\simeq 0.18$ fm.
Renormalization constants for the naive and improved vector current
 and  the naive axial current are computed
using Ward identities. The renormalization factors are very close to unity,
and the spectroscopy of light hadrons and the pseudoscalar and vector
decay constants agree well with simulations at
smaller lattice spacings (and with experiment).
\endabstract

\pacs{COLO-HEP-421}

\section{Introduction}
This paper is a continuation of earlier work \cite{HYP1,FATCLOVER} studying the
properties of  a particular type of improved action for studies
of quenched QCD, with fermion-gauge field couplings parameterized by
``fat links,'' and a lattice anomalous magnetic moment ``clover''
term. One of the actions studied here has fermionic couplings extending over
a hypercube. The other action is the fat link clover action.
 Both actions are $O(a^2)$ improved. Features related to 
their chiral properties appear to be superior to those of standard
discretizations,
 though this does not correspond to a realization of an exact lattice symmetry
relation.
The goal of this paper is to compare a particular implementation
 of a fixed point
(FP) action for quenched QCD in four dimensions to
 a simpler improved action, via
the usual tests of improvement. Much more elegant studies have been
carried out in two dimensions \cite{TWODIM}.
On the way,  the fat link clover action turns out to be an attractive
alternative.

The hypercubic 
action is  inspired by the fixed point action program \cite{FP},
applied to fermion actions for QCD in
 four dimensions\cite{Wiese,BW,FPF,MITFP1,MITFP2}.  FP actions are
classically perfect, which means that they have the following
 desirable properties: 

First, their spectrum has no lattice spacing $a$-dependent corrections
of the form $a^n$  for any $n$.
Second, and probably more importantly, FP actions satisfy
the Ginsparg-Wilson \cite{GW} remnant chiral symmetry condition,
namely that the anticommutator of the propagator with $\gamma_5$ is
a local operator.
As a result, they  suffer no additive quark mass renormalization and
no multiplicative renormalization of axial currents
\cite{PHGW}, and satisfy the index theorem \cite{PLN,LUSCHER}.
It would clearly be a desirable thing to have a version of such an
 action which could be used in numerical simulations, and this paper describes
a candidate action which seems to satisfy all of the properties of an
FP action very well, though not exactly.

The particular hypercubic  action I will test is one 
whose free-field
limit is constructed by blocking out of the continuum. It has the
usual clover term, normalized to its tree-level value as described in
\cite{HYP1}.
The gauge connections of both the hypercubic action and clover action
I study are replaced by APE-blocked \cite{APEBlock}
 links,
\beea
V^{(n)}_\mu(x) = &
(1-\alpha)V^{(n-1)}_\mu(x) \nonumber  \\
& +   \alpha/6 \sum_{\nu \ne \mu}
(V^{(n-1)}_\nu(x)V^{(n-1)}_\mu(x+\hat \nu)V^{(n-1)}_\nu(x+\hat \mu)^\dagger
\nonumber  \\
& +  V^{(n-1)}_\nu(x- \hat \nu)^\dagger
 V^{(n-1)}_\mu(x- \hat \nu)V^{(n-1)}_\nu(x - \hat \nu +\hat \mu) ),
\label{APE}
\eea
with  $V^{(n)}_\mu(x)$  projected back onto $SU(3)$ after each step, and
 $V^{(0)}_\mu(n)=U_\mu(n)$ the original link variable. 
I take $\alpha=0.45$ and $N=10$ smearing steps,
chosen because this fattening produces a very small additive mass
renormalization.

There are two differences between this work and Ref. \cite{HYP1}.
The first involves physics: since the completion of that work
 we \cite{FATCLOVER} have
discovered that the properties of lattice QCD (specifically the
spectrum of real fermion eigenmodes) changes as the lattice spacing coarsens,
and that above about $a_{max}\simeq 0.2$ fm the would-be zero modes cannot be
separated from the doubler modes. Since there is a qualitative change in 
the underlying dynamics,
it does not seem to be appropriate to attempt to extrapolate to
the continuum physics which
involves chiral symmetry from lattice spacings greater than $a_{max}$.
The second difference is that in this paper I study FP operators,
the analogs of the familiar ``rotated'' operators of the Symanzik program.
For the action used in \cite{HYP1}, the recursion relation for the FP operators
is contaminated by a  redundant eigenvector. I chose to find a different
renormalization group transformation (blocking out of the continuum with
a Gaussian blocking function)
and a different hypercube action, the ``Gaussian'' hypercubic action,
 in order to construct the operators.

The outline of the paper is as follows:
After reviewing the properties of FP actions, I discuss the ways that
approximations to FP actions are imperfect. I summarize the physics 
features of fat link actions. Then I present tests of
spectroscopy and of vector and axial current matrix elements, at
lattice spacing $a=0.18$ fm. I end with some conclusions and speculations.

\section{A Quick Review of Fixed-Point Action Formalism}

To find a FP action for QCD, one begins with a set of fermionic ($\psi_n$,
$\bar \psi_n$) and gauge field  ($U_\mu(n)$) variables defined either on a fine
lattice or  in the continuum.
The fine action is defined as
\begin{equation}
S = \beta S_g(U) + \bar\psi_i \Delta(U)_{ij}\psi_j  .
\end{equation}
$S_g$ is the gauge action, $i$, $j$ label sites, and $\Delta(U)$ is the
fermion action. Introducing a renormalization group (RG)
blocking kernel with a pure gauge piece $T_g$ and a
fermionic piece parameterized by a constant $K$ and
a blocking function $\alpha_{n_b,n}$ normalized by $b$
\begin{equation}
T =  \beta T_g(U,V) +
K \sum_{n_b}(\bar \Psi_{n_b} - \bar\psi_n( b \alpha_{n,n_b}^\dagger ))
(\Psi_{n_b} -( b \alpha_{n_b,n'})\psi_{n'}),
\end{equation}
one integrates out
 the fine degrees of freedom to  construct an action involving
 coarse-grained
variables $\Psi_{n_b}$, $\bar \Psi_{n_b}$ and $V_\mu(n_b)$.

The renormalization group equation
\begin{equation}
{\rm e}^{-S'} = \int d\psi d\bar\psi dU {\rm e}^{-(T+S)}
\end{equation}
has a pure gauge FP at $g^2=0$ ($\beta \rightarrow \infty$).
 In that limit the
gauge action dominates the integral; its RG equation is given by the same
steepest-descent equation as for  a pure gauge model
\begin{equation}
S^{FP}(V)=\min_{ \{U\} } \left( S^{FP}(U) +T(U,V)\right),
\label{RGG}
\end{equation}
while the fermions sit in the gauge-field background. The fermion action
remains quadratic in the field variables, and the transformation of
the fermion action is given most easily in terms of the propagator
\begin{equation}
(\Delta'(V))^{-1}_{n_b,n'_b} =   {1\over K} \delta_{n_b,n'_b}
+  b^2 \alpha(U)_{n_b,n} (\Delta(U))^{-1}_{n,n'}\alpha(U)_{n',n'_b}^T
\label{RGF}
\end{equation}
where $\{U\}$ is the field configuration which minimizes Eqn.
\ref{RGG} for a given $\{V\}$. 
For a blocking factor $F$, it is useful
to rescale $b \alpha_{n_b,n} = F^{3/2}\Omega_{n_b,n}$
with the Fourier transform of $\Omega_{n_b,n}$ for a free theory normalized
to $\Omega(q) \simeq 1 + O(q^2)$. Then (again in free field theory)
 the momentum-space FP equation
for the propagator becomes
\bee
(\Delta'(q))^{-1} = {1\over K} + {1 \over F}\sum_l
|\Omega({{q+2\pi l}\over F})|^2 \Delta({{q+2\pi l}\over F})^{-1}.
\label{FPACTION}
\ee

Another useful quantity is the minimizing field, the value of the fine
field (as a function of the coarse field) which minimizes the exponential in
 the Gaussian fermion integral:
\beea
\psi_{min}(n)  & = & \zeta(n-Fn_b)\Psi(n_b) \nonumber \\
       & = & \sum_{m,m_b} \Delta^{-1}(n-m)b \alpha_{m,Fm_b}
 \Delta(m_b-n_b) \Psi(n_b) \nonumber \\
\label{FPFIELD}
\eea

A related object is the fixed-point field \cite{FP,FP2}, a local average
 of fields which blocks into itself under a RG transformation,
\bee
\psi_{FP}(n)  =\sum_m w(n-m)\psi(m) .
\ee
The averaging function obeys the recursion relation
\bee
\sum_m w(Fn_b-m) \zeta(m-Fm_b)= \lambda w(n_b-m_b)
\label{OOPS}
\ee
with $\lambda = 1/F^{3/2}$ carrying the canonical dimension of a fermion field.
As the scale of the blocking transformation $F$ goes to infinity,
the FP field and minimizing field coincide. Correlators involving
the FP field have no power-law cutoff effects--they are also classically
perfect. Thus they play the same role for a FP action
 as the familiar ``rotated operator'' does
in the Symanzik improvement program.

The easiest way to find the FP field, for an action defined by
an RG transformation with a scale factor F, is to solve Eqn. \ref{OOPS}
self-consistently for $w(n)$.  Unfortunately, for the action studied
in Ref. \cite{HYP1}, with a blocking factor $F=2$ RGT,
 $\lambda= 1/F^{3/2}$ is not the leading eigenvalue:
The leading eigenvector is $w(n)=\Delta(n)$, with an eigenvalue of about
$2.56/2^{3/2}$.
The phenomenon has been described by P. Kunszt in \cite{PKUNSZT}.
In terms of physics this does nothing: $w(n) = \Delta(n)$ just contributes
a contact term to correlation functions. However, from an engineering
point of view  is a serious problem, because it is quite difficult to 
determine the FP field by solving the recursive equation for $w$, and
pulling out a nonleading eigenvector.

For a free field action constructed by blocking out of the continuum
($F\rightarrow\infty$ or equivalently, $m$ a continuous variable in
the background of fixed $n_b$), the free FP field is readily constructed
using the momentum-space version of Eqn. \ref{FPFIELD}.
Therefore, because I wanted to study FP operators, I decided to
construct a new action by following the MIT group \cite{MITFP1,MITFP2}
and blocking directly
out of the continuum. There is a price associated with this--I can
no longer solve FP equations for the propagator in a nontrivial gauge
field background, as I did in \cite{HYP1}.
I will follow a hybrid approach, of constructing the free field FP
action, making it gauge invariant by using connections made of fat links,
and just tuning the fattening to optimize the chiral properties of the action.
The construction of the free action is described in the Appendix.

\section{Imperfection}
\subsection{Comparison to the standard tests of imperfection}
An FP action is classically perfect, but an approximation to a FP
action is not. It is thus an interesting exercise to ask how imperfect
an approximate FP action will be.

One can approximately reconstruct the free field FP action and field
by blocking out of the continuum. In Eqn. \ref{FPACTION} the $l=0$
term in the mode sum is the most singular for small $q$
 and the next higher order
term is the $1/K$ term. Keeping only those contributions,
a few lines of algebra yields
\bee
\Delta(q) = {1 \over \Omega(q)^2}
 {{-i \gamma\cdot q +m + {{q^2+m^2}\over{K\Omega^2}}}
\over{1+{{2m}\over{K\Omega^2}} +{{q^2+m^2}\over{(K\Omega^2)^2}} }}
\label{MYACT}
\ee
\bee
\zeta(q) = {1 \over \Omega(q)^2}
 {{1 + {{i \gamma\cdot  q + m}\over{K\Omega^2}} }
\over{1+{{2m}\over{K\Omega^2}} +{{q^2+m^2}\over{(K\Omega^2)^2}} }}
\label{MYROT}
\ee
As $\Omega \sim 1+ q^2 + \dots$, the reader will recognize these formulas
as rather baroque variations on the usual Symanzik-improved action
and rotated field -- up to an overall normalization constant.
Indeed
\bee
\Delta \simeq ( \gamma\cdot D + m - {1\over K}(D^2-m^2))(1 - {{2m}\over K}))
+ \dots
\ee
\bee
\zeta \simeq (1 + {{m - \gamma\cdot D )}\over K})(1 - {{2m}\over K})
+ \dots
\ee
and so we identify the Wilson $r-$parameter with $1/K=ra/2$.

The most serious lattice artifacts of a Wilson-like action come from
its vertices. In the absence of an explicit construction of a FP action
in a nontrivial background gauge field configuration, what can one say?
First of all, a simple expansion of the action in powers of $a$ shows that
any hypercubic action made of thin or fat links has $O(a)$ contributions to
 its vertex, just like the Wilson action. These contributions must cancel in
perturbation theory 
between the scalar vertices and an additional anomalous magnetic
moment ``clover'' term, which of course is absent in the free theory.

In principle, the clover term could have any coefficient, but 
one fact we know is that the spectrum of a FP action is classically perfect.
One can consider the spectrum of a fermion with infinitesimal momentum $p$
in an infinitesimal external
magnetic field $B$, $E=m_0 +p^2/2m_K - B/m_B+\dots$ and constrain
 $m_B=m_K=m_0$, by
hand, if necessary.
This constraint turns out to be identical to the requirement that the clover
term have its usual tree level value.
Now the calculation just paraphrases the old perturbative result
of Heatlie, et.~al. \cite{HEATLIE} and the result is as expected:
the spectrum of a hypercubic approximation to a FP action,
with the clover term, is improved through $O(a^2)$,
and if matrix elements are measured using FP operators, they
also have no $O(a)$ discretization errors.

\subsection{Chiral Properties}
Violations of chiral symmetry in Wilson-like actions
 are an old story \cite{CHIRALWI}:
Write the fermion action as
\bee
\Delta(m)=\Delta_0 + m_0
\ee
and consider a flavor nonsinglet chiral rotation
 $\psi(x) \rightarrow 
\exp (i \epsilon^a(x){{\lambda^a}\over 2} \gamma_5)\psi(x)$,
 $\bar \psi(x) \rightarrow \bar\psi(x)\exp
 (i \epsilon^a(x){{\lambda^a}\over 2} \gamma_5)$.
The Ward identity for this rotation is
\beea
{{\delta \langle O(x_1,\dots,x_n) \rangle}\over {\delta \epsilon^a(x)}}
& = & \langle  O(x_1,\dots,x_n)\partial_\mu J_\mu^5(x)^a \rangle\nonumber \\
& & -\langle  O(x_1,\dots,x_n) m_0 \bar \psi(x){{\lambda^a}\over 2} \gamma_5
 \psi(x)\rangle \nonumber \\
& & -\langle  O(x_1,\dots,x_n)
(\bar \psi(x){{\lambda^a}\over 2}\{ \gamma_5,\Delta_0\}
 \psi(x))\rangle. \nonumber \\
\label{WARD1}
\eea
The last term is the lattice artifact. In a generic non-FP action, it will mix
with lower-dimensional operators 
\bee
\bar \psi(x)\{ \gamma_5,\Delta_0\} \psi(x)= (1-Z_A)\partial_\mu J_\mu^5(x)
+ \Delta m \psi(x) \gamma_5 \psi(x) + O(a)+\dots
\label{WARD2}
\ee
to give a nonzero additive renormalization to the quark mass and an
overall multiplicative renormalization to the axial current. It
is also responsible for mixing of opposite parity operators in 
matrix elements appropriate to, for example,
 $B_K$. In perturbation theory all of
these mixings are due to one-loop graphs and are $O(g^2)$.

The Ginsparg-Wilson relation eliminates these mixings and renormalizations
\cite{PLN} (or to be more precise converts them into local  corrections
to the Ward identities), by canceling the propagator connecting the operator
to the current
with the anticommutator against the action term from the anticommutator.

The  ingredient of an approximate FP action most necessary to reproduce
this feature of a FP action is a fat link. The dominant graphs contributing
to $Z_A-1$ and $\Delta m$ contain tadpoles, whose contributions from
large $q^2 $are
suppressed by the soft $q \bar q g$ vertices of a fat link action
\cite{FATCLOVER}. The presence of the fat link in the one explicit
realization of the Ginsparg-Wilson relation, the
 ``overlap action'' \cite{OVER}
\bee
\Delta =
 {K\over 2}( 1 - (1-{2\over K} \Delta_0)/\sqrt{| (1-{2\over K} \Delta_0|^2}),
\ee
using a thin link $\Delta_0$,
can be seen easily by making a hopping parameter expansion in $\Delta_0$.

Finally, can one quantify the size of chiral violations expected for
an approximate FP action?
Imagine constructing a FP action by beginning with the Wilson action
and performing a series of factor-of-two RGT's. The Wilson term violates
the Ginsparg-Wilson relation. It is a dimension-five operator, and so under
each blocking step its size decreases by a factor of $1/2$. After $N$ blocking
steps the action will have a range $O(N)$ and the size of the
 violating operator will
be $O(\exp(-N \ln 2)$. This exponential decrease of chiral violations with
range of the action seems superficially to be the same behavior as is
seen with domain wall fermions (where $N$ is the length of the fifth
dimension \cite{BLUM}).
One can easily see this behavior in explicit calculations using free field
theory.

\section{Tuning for chiral behavior}
The most visible aspect of bad chiral behavior 
 for Wilson-like fermions is 
 the presence of exceptional configurations, which arise when the
 Dirac operator $\Delta_0$
has a real eigenmode at  $\lambda=-m$, a value 
which happens to coincide with the bare mass $m$ at which one attempts
to construct a propagator for $\Delta_0+m$ \cite{FNALEX}
We have argued in Ref. \cite{FATCLOVER} that an action whose
gauge connections are fat links has a narrower range of real eigenvalues
than a thin link action does.

\begin{figure}[thb]
\epsfxsize=0.8 \hsize
\epsffile{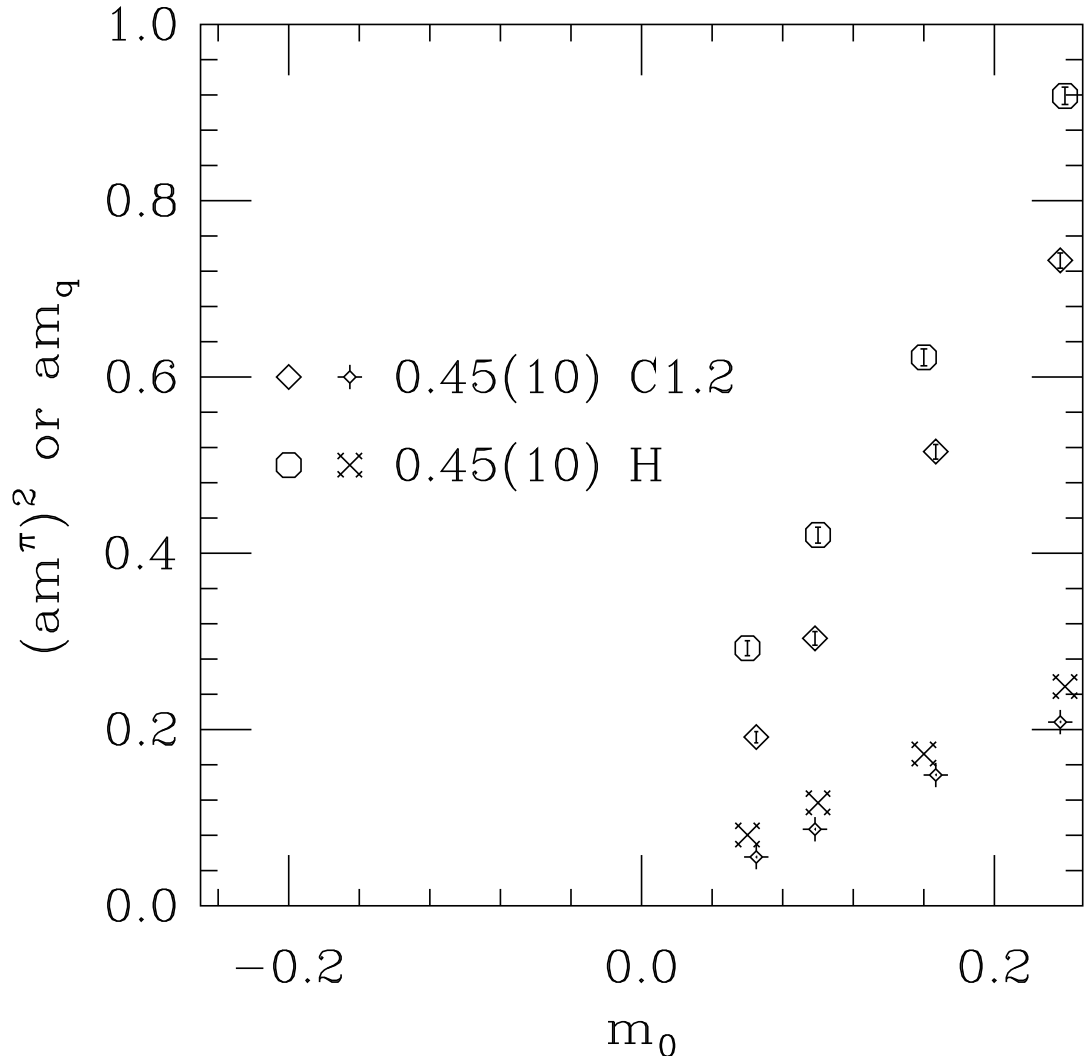}
\caption{
$m_\pi^2$ and the quark mass vs. bare quark mass for various fermion actions.
Data are diamonds and fancy diamonds for
$m_\pi^2$ and the quark mass
from the PCAC relation respectively, for the
fat link clover action of Ref. 2, with the clover term boosted
by a factor 1.2, and octagons and fancy crosses for the Gaussian hypercubic
action.
Both actions have fat links with
$\alpha=0.45$, $N=10$.
}
\label{fig:pisqb}
\end{figure}

In Ref. \cite{FATCLOVER} we proposed a method for tuning an action
to optimize its chiral behavior. This involved measuring the spread of real
eigenmodes of the Dirac operator, and varying the amount of fattening
and the size of the clover term to narrow the spread of real eigenmodes.
This is computationally costly and if the amount of fattening is large,
produces only the unsurprising result, that a clover coefficient close
to the tree-level value is the optimum one.

Here, I take a simpler (though more tasteless) approach: fix the
clover coefficient to its tree-level value and vary the amount of fattening
until the additive mass renormalization, measured naively by extrapolating
the squared pion mass linearly to zero,
appears to be small. This is done at small quark masses, but not at masses
which are so small that the quenched approximation breaks down due
to the presence of unpaired instanton modes \cite{QBREAK}.
The results are shown in Fig. \ref{fig:pisqb}.
Data are diamonds and fancy diamonds for $m_\pi^2$ and the quark mass
from the PCAC relation respectively, for the
fat link clover action of Ref. \cite{FATCLOVER}, with the clover term boosted
by a factor 1.2, and octagons and fancy crosses for the Gaussian hypercubic
action.
Both actions have fat links with
$\alpha=0.45$, $N=10$.
A jacknife fit of the four lightest masses
of $m_\pi^2=A(m_0 - m_c)$ or $m_q = A(m_0 - m_c)$
(this is the lattice quark mass from the PCAC relation; see Sec. VI.C)
gives $m_c = -0.0226(15)$ from $m_\pi^2$ and
$m_c=-0.0248(7)$ from the quark mass, for the hypercubic
action, and $m_c=0.0023(19)$ and -0.0002(18) for the fat link clover
action.

\begin{figure}[thb]
\epsfxsize=0.8 \hsize
\epsffile{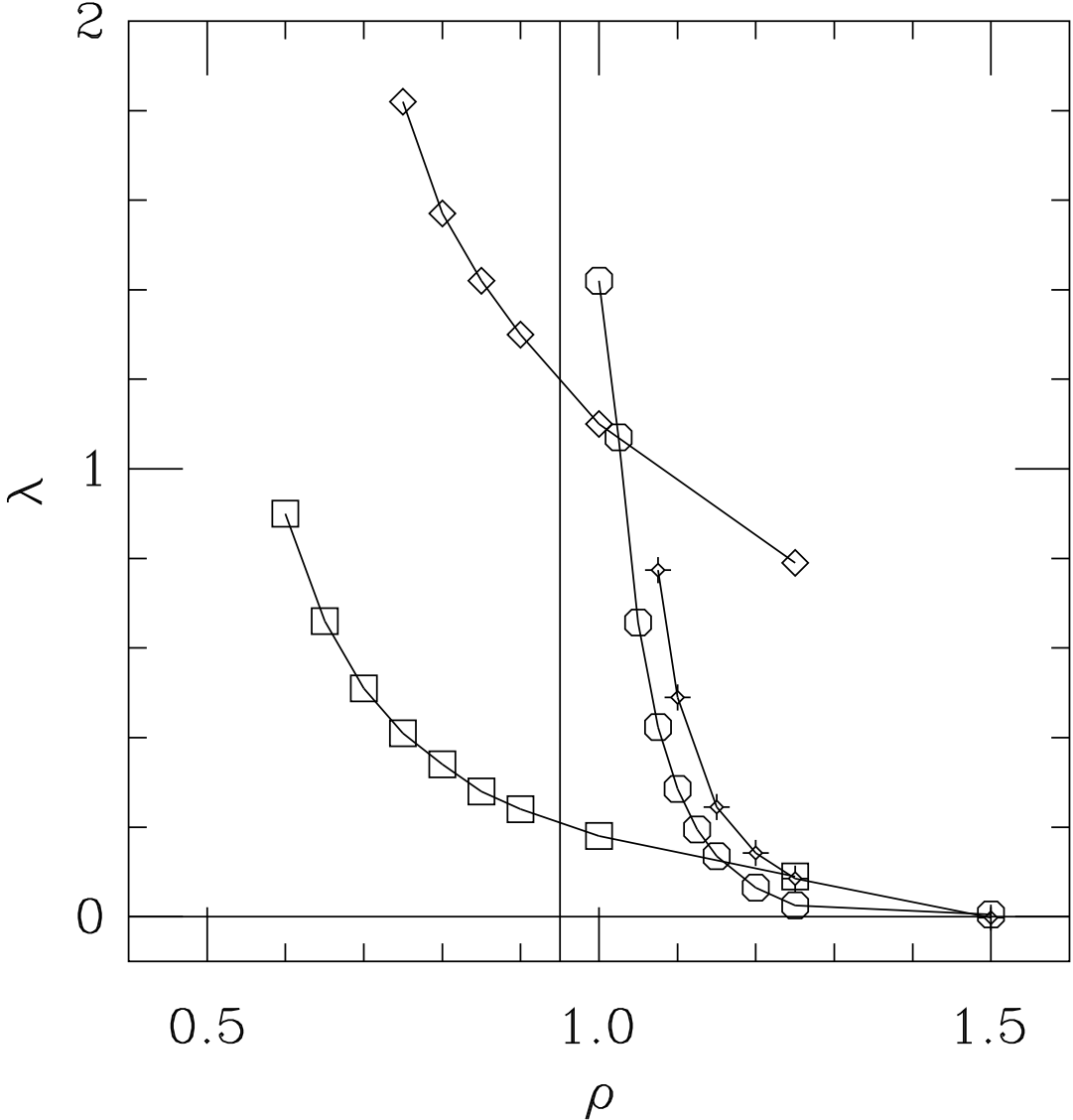}
\caption{
Smallest real eigenmode of the massless Dirac operator
 vs. instanton size for various fermion actions.
Diamonds and squares show the usual thin-link Wilson and clover actions
while the
fancy diamonds
 show the fat link clover action and octagons show the hypercubic
 action (both with fattening $\alpha=0.45$ and $N=10$ steps).
}
\label{fig:mvsrho}
\end{figure}

To make a controlled comparison of the zero mode
spectra of these actions, I show in Fig. \ref{fig:mvsrho}
 the real eigenmode spectrum of several of these
actions
on a family of single instanton configurations.
The construction of the instantons is described in Ref. \cite{INST}.
They are $SU(2)$ instantons in singular gauge centered at
$(L/2+1/2,L/2+1/2,L/2+1/2,L/2+1/2)$ in a periodic lattice of size $L=8$.
The vertical line at $\rho \simeq 0.95$ marks the smallest radius
visible to the approximate FP  gauge action of Ref. \cite{HYP1}.
Diamonds and squares show the usual thin-link Wilson and clover actions
while the other plotting symbols show fat link actions. As the
 instantons shrink, the real eigenmodes become more positive, and then,
when the instanton disappears,
they collide with another real eigenmode (a doubler) to produce a
complex conjugate pair of eigenvalues.

For future reference, recall that an action obeying the
 Ginsparg-Wilson relation would have a zero eigenmode on a topologically
nontrivial configuration, but when the instanton ``falls through the lattice''
the eigenmode should jump discontinuously to some positive value.
Presumably the more improved an action is, the sharper the rise of the
real eigenmode as the instanton size shrinks. All  fat link
 actions, both hypercube and clover,
 satisfy that criterion. Based on our studies with
fat link clover actions\cite{FATCLOVER},
 we believe that this feature, like the small
mass renormalization, is caused by the fat link, not by features
of the fermion action which could be seen in the free-field limit.

\section{Numerical Tests}
\subsection{Spectroscopy}
I carried out quenched spectroscopy on a set of 80 $8^3\times 24$
 background gauge
field configurations with a lattice spacing $a\simeq 0.18$ fm
using the approximate FP gauge action of Ref. \cite{HYP1}, at
$\beta=3.70$. As a fiducial test of a simpler action I recomputed
the spectrum of the fat link clover action of Ref. \cite{FATCLOVER}
on the same set of background configurations (in order to remove
all effects of the gauge fields from comparisons). This action also
has $\alpha=0.45$, $N=10$ APE-smeared fat links and its clover term
is rescaled by a factor $C_{SW}=1.2$.

The cost of the hypercubic action compared to the clover action
 is a factor of about 17 per
multiplication $\psi = (\Delta + m)\chi$ during the iterative
construction of the propagators. Both the fat link clover action and
 hypercubic
actions appear to require about half the number of iterations as the
usual thin link clover action
to converge to the same residue $|\psi-(\Delta+m)\chi|^2/|\chi|^2$,
presumably because their high Fourier modes decouple from the gauge fields.
Neither of these actions developed any exceptional configurations over the
studied range ($\pi/\rho \ge 0.54$ and 0.64).

 The spectrum analysis is completely standard\cite{HYP1}.
Both actions reproduce the results of spectroscopy calculations of
more conventional actions at smaller lattice spacings.
The spectroscopy is tabulated in Tables \ref{tab:g370} and \ref{tab:flc370}.

\begin{table}
\begin{tabular}{|c|l|l|l|l|}
\hline
$am_q$ & PS  & V  &  N  &  $\Delta$ \\
\hline   0.32 &  1.116( 3) &  1.246( 5)&  1.935(11) &  2.040(12)  \\
   0.24 &  0.959( 5) &  1.117( 6)&  1.726(10) &  1.862(16)  \\
   0.16 &  0.789( 6) &  0.994( 8)&  1.517(13) &  1.735(17)  \\
   0.10 &  0.649( 6) &  0.902(13)&  1.370(15) &  1.622(23)  \\
   0.06 &  0.541( 7) &  0.851(19)&  1.264(21) &  1.543(29)\\
\hline
\end{tabular}
\caption{Table of best-fit masses,
 Gaussian hypercube action, $\beta=3.70$ ($aT_c=1/4)$.}
\label{tab:g370}
\end{table}

\begin{table}
\begin{tabular}{|c|l|l|l|l|}
\hline
$\kappa$ & PS  & V  &  N  &  $\Delta$ \\
\hline
   0.114 &  1.094( 3) &  1.226( 4)&  1.887(10) &  1.984( 9)  \\
   0.116 &  0.977( 3) &  1.137( 5)&  1.741( 9) &  1.862(12)  \\
   0.118 &  0.856( 5) &  1.048( 6)&  1.595(10) &  1.740(16)  \\
   0.120 &  0.718( 5) &  0.960( 9)&  1.449(12) &  1.616(25)  \\
   0.122 &  0.551( 7) &  0.866(18)&  1.280(19) &  1.555(25) \\ 
   0.123 &  0.437( 7) &  0.807(28)&  1.190(30) &  1.482(32)\\
\hline
\end{tabular}
\caption{Table of best-fit masses, fat link clover action,
 $\beta=3.70$ ($aT_c=1/4)$.}
\label{tab:flc370}
\end{table}

\begin{figure}[thb]
\epsfxsize=0.8 \hsize
\epsffile{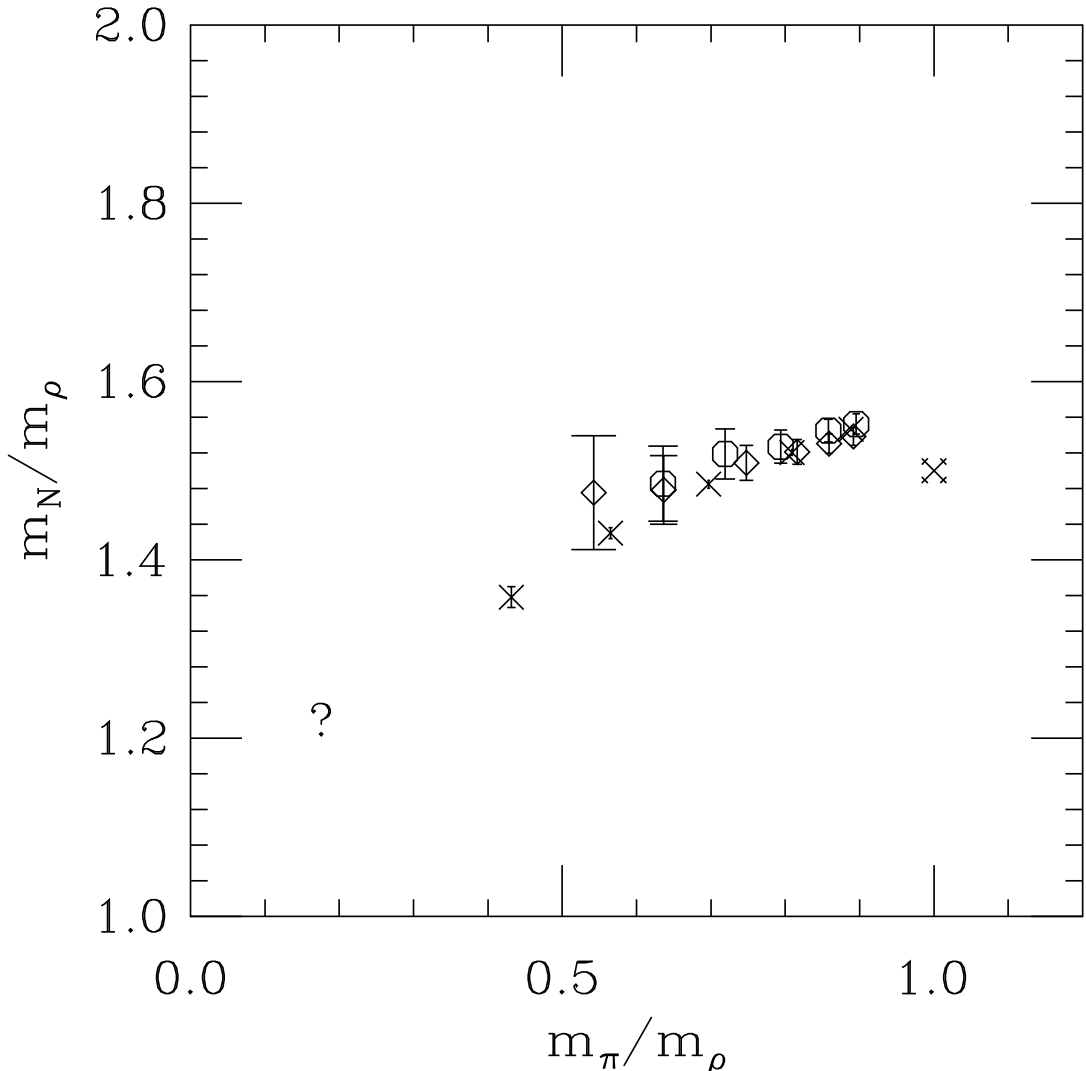}
\caption{
Edinburgh plot showing $\beta=6.15$ staggered fermions from Ref. 25 (crosses),
the fat link clover action (diamonds) and the Gaussian hypercube action
(octagons).
}
\label{fig:ed370}
\end{figure}

\begin{figure}[thb]
\epsfxsize=0.8 \hsize
\epsffile{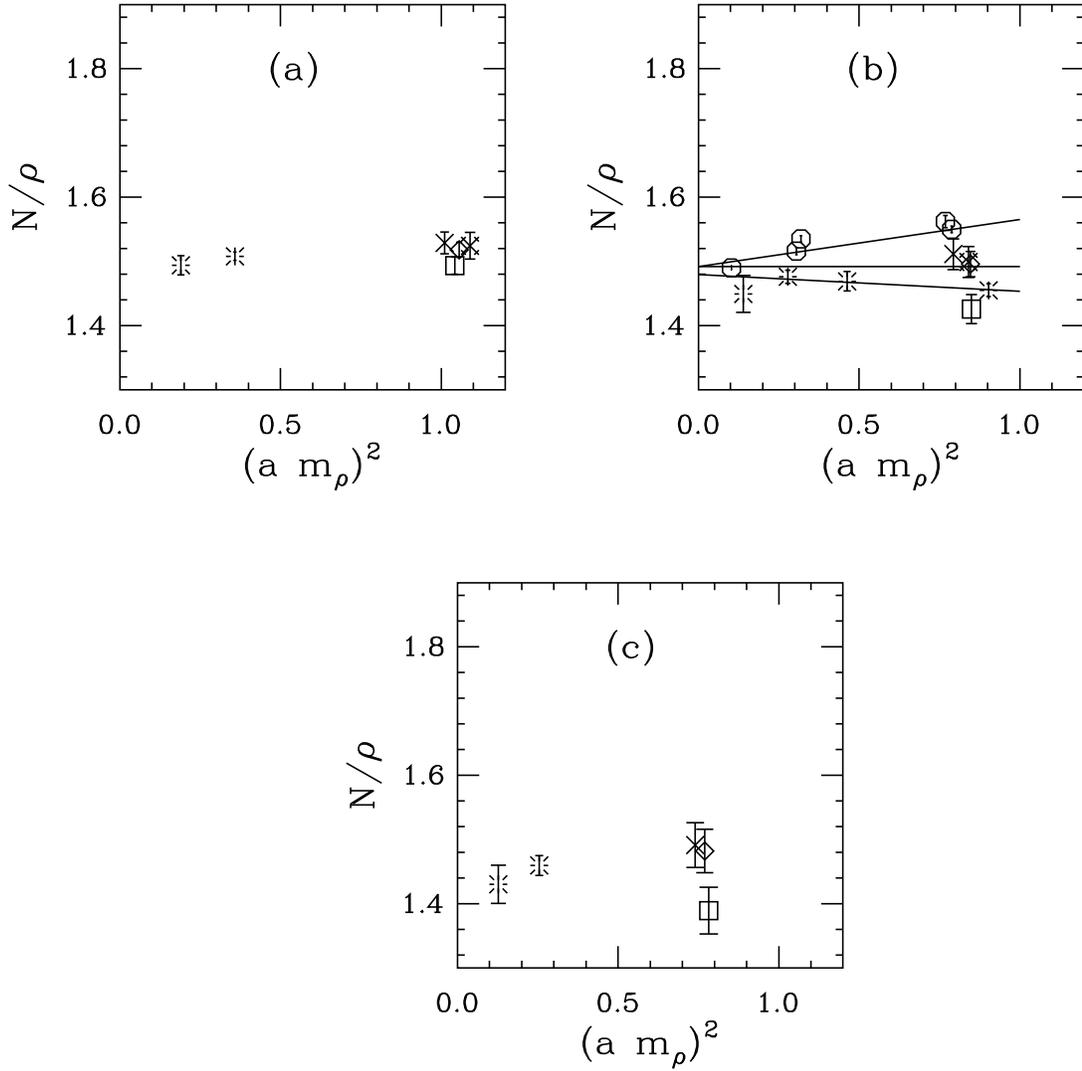}
\caption{
$N/\rho$ mass ratio interpolated to
$\pi/\rho=0.80$ (a), 0.70 (b), 0.65 (c)
the
nonperturbatively improved (thin-link) clover action (bursts),
$c_{sw}=1.2$ fat link clover action (squares on $\beta=5.7$ Wilson gauge
actions, diamonds on FP gauge configurations), hypercubic action
of Ref. 1 (fancy crosses), and Gaussian hypercubic action (crosses).}
\label{fig:allrat}
\end{figure}

To show the data graphically,
I begin with an Edinburgh plot, Fig. \ref{fig:ed370}, containing
$\beta=6.15$ staggered fermion data \cite{MILCKS} and data from the two
actions shown here, the fat link clover action and the Gaussian hypercubic
action. I compare scaling violations in hyperfine splittings by
interpolating the data to fixed $\pi/\rho$ mass ratios and plotting
the $N/\rho$ mass ratio vs. $m_\rho a$. I do this at three
$\pi/\rho$ mass ratios, 0.80, 0.70, and 0.65, in Fig. \ref{fig:allrat}.
The bursts are from the nonperturbatively improved (thin link)
 clover action of Refs.
\cite{ALPHA}  and \cite{SCRI}.
The octagons show data from staggered fermions at $\beta=6.15$.
The other plotting symbols show our test actions:
 They are the hypercubic 
action of Ref. \cite{HYP1} (fancy crosses) and of this work (crosses)
and the $c_{sw}=1.2$ fat link clover
 action (squares on $\beta=5.7$ Wilson gauge
actions \cite{FATCLOVER}, diamonds on FP gauge configurations).
The data indicate that the improved kinetic properties of the
hypercubic action do not affect hyperfine splittings very strongly.
The two outer straight lines in the figure are linear fits in $(am\rho)^2$
to the staggered and nonperturbatively improved clover data.
The inner straight line is a constant at the value of the extrapolated
$a=0$ value of the staggered data.

The fat link clover action on Wilson gauge configurations has a slightly
larger hyperfine splitting than the other actions. Could this be
an effect of the gauge action on the spectrum? Could it be due to the fact that
the Wilson gauge action overproduces small instantons compared to the
FP gauge action?

\begin{figure}[thb]
\epsfxsize=0.8 \hsize
\epsffile{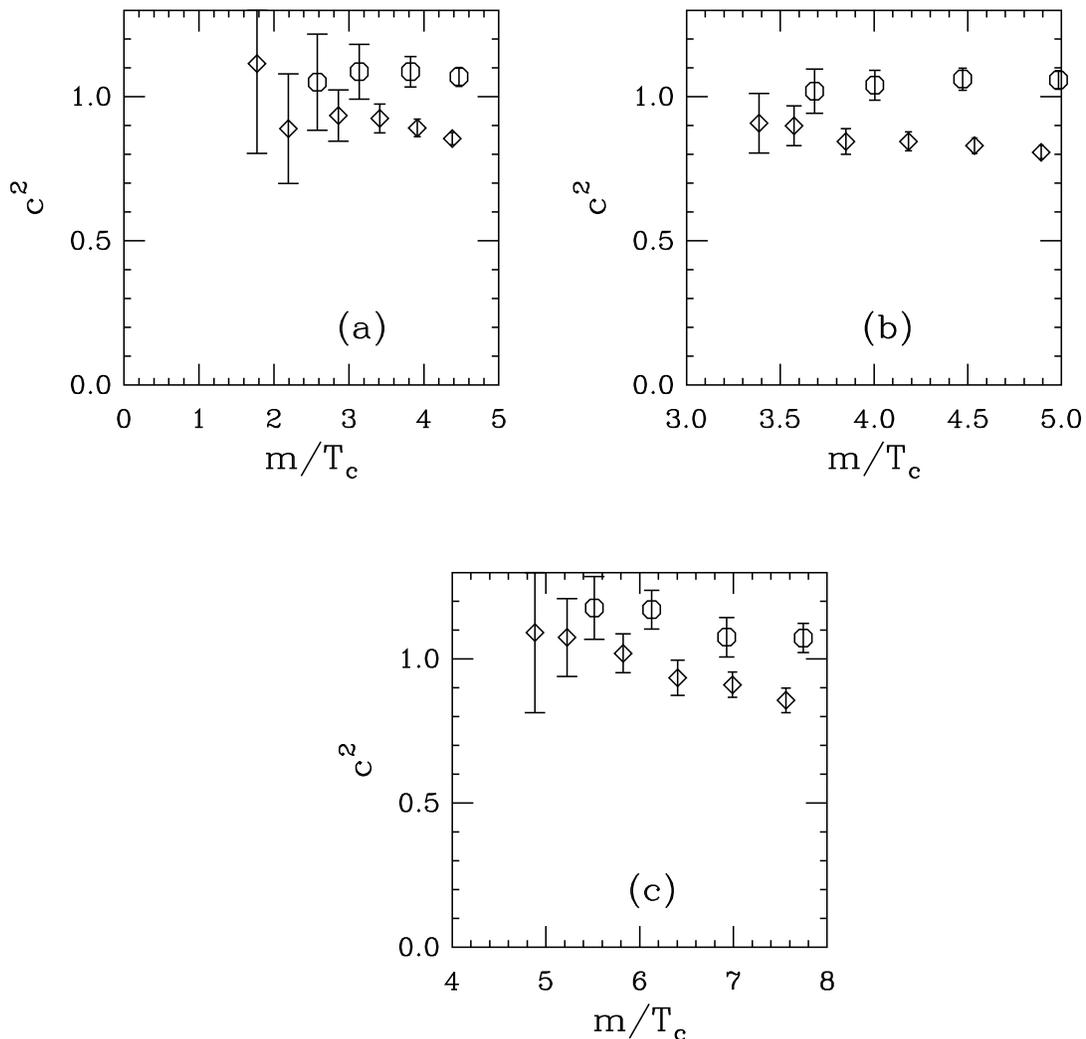}
\caption{$c^2$ from Eqn. 30 for the fat link clover action
(diamonds) and Gaussian hypercubic action (octagons),
for (a) pseudoscalar (b) vector and (c) nucleon correlators.
Hadron masses are shown in units of the deconfinement temperature
 $T_c=1/(4a)$.}
\label{fig:c2comp}
\end{figure}

Differences appear when one compares the
dispersion relations (Fig. \ref{fig:c2comp}). The
squared speed of light,
$c^2 = (E(p)^2-m^2)/p^2$, for $\vec p= 2\pi/8(1,0,0)$, is computed
by performing
a correlated fit to the two propagators. At larger mass the
clover action's $c^2$ falls away from unity while the hypercubic action
shows a fairly constant $c^2-1 \simeq 0.025$.
To achieve the same mismatch of $c^2-1$ as the hypercubic action,
at the same $m/T_c$, one
would have to decrease the lattice spacing by about a factor of
roughly two (assuming $c^2-1 \simeq O(a^2)$).

\subsection{Vector Currents}
There are many definitions of conserved currents. The simplest one
is constructed
by writing \cite{GW,PLN}
$U_\mu(n)=1+igA_\mu(n)+\dots$
and defining
\bee
J_\mu(n)= -i {{\delta S}\over{\delta A_\mu(n)}}
\label{THIN}
\ee
In these actions
I have replaced the thin links by fat links,
$V_\mu(n)=1+igB_\mu(n)+\dots$
which in linear approximation corresponds to
\bee
B_\mu(n)= \sum_{r,\nu}c_{\mu\nu}(r)A_\nu(r)
\ee
for some smearing function $c_{\mu\nu}(r)$. It is more
convenient to replace Eqn. \ref{THIN} with a  ``fat conserved current''
\bee
\hat J_\mu(n)= -i {{\delta S}\over{\delta B_\mu(n)}}.
\label{FATJ}
\ee
Are $\partial_\mu J_\mu(n)=0$ and $\partial_\mu \hat J_\mu(n)=0$
consistent?  Yes.
In general, $c_{\mu\nu}$ has two parts. The first is
a term proportional to $\delta_{\mu\nu}$, which is parity-even,
($\delta_{\mu\nu}c_S(r)A_\mu(r)$ with
$c_S(r+\hat\lambda)=c_S(r-\hat\lambda)$),
and a term which is proportional to $(1-\delta_{\mu\nu})$,
$(1-\delta_{\mu\nu})c_A(r)A_\nu(r)$ with
$c_A(r+\hat\lambda)= -c_A(r-\hat\lambda)$.
The ``1'' terms in $1-\delta_{\mu\nu}$ are of the form
$\sum_{r,\nu}c_A(r)(J_\nu(n+r)-J_\nu(n+r-\nu))$ which vanishes if
$J_\mu$ is conserved, and so $\hat J_\mu(n)=\sum_r d(r)J_\mu(n+r)$
for some function $d(r)$.
Thus conservation of the usual current implies (all rather trivially)
conservation of the  associated fat current.

In what follows, the conserved vector current is defined using
Eqn. \ref{FATJ}. The hypercubic action has a ``conventional'' normalization
$S = m \bar \psi \psi + \dots$, in contrast to the ``kappa'' normalization
of the clover action $S = \bar \psi \psi - \kappa \bar\psi D \psi $.
 In either case, the local current is
 $\bar \psi \gamma_\mu \psi$ and
the improved current is $\bar{(\zeta \psi )} \gamma_\mu(\zeta \psi)$.

Now  recall the classification of definitions of the current\cite{MSV}:
\bee
\langle f | J_\mu |i \rangle_{cont} =
 Z_{J^X}\langle f | J_\mu^X |i \rangle_{latt} + O(a) + O(g^2 a) +\dots
\label{CVC}
\ee
where $Z_{J^X} = F(m)( 1 + c_1 g^2 + \dots)$ (introducing the field
rescaling of the usual Wilson or clover actions).
If the current is conserved (C), its $Z=1$. If it is improved (I), the
$O(a)$ term is zero. The  conserved current of either action is not improved.
The ''rotated'' or FP currents are improved but not conserved.
It's conventional to rescale the results of the clover action so that
$F(m=0)=1$.
This amounts to rescaling the clover field
 $\psi \rightarrow \sqrt{2\kappa}\psi$.
 I will present all results that way even though it is unnecessary
for this method of finding renormalization factors.

I measured three kinds of matrix elements:
\bee
R^{X}_{\pi}={{\langle \pi(T) | \pi(0) \rangle}
 \over
 {\langle \pi(T)| J_0(t)^X|\pi(0) \rangle}},
\label{FWD}
\ee
\bee
R^{X,Y}_{\pi,\vec q}={{\langle \pi(T,0)| J_0(t,\vec q)^X| \pi(0,-\vec q)
 \rangle}
 \over
 {\langle \pi(T,0)| J_0(t,\vec q)^Y|\pi(0, \vec q) \rangle}},
\label{RPI}
\ee
and the ratio
\bee
R^{X,Y}_\rho(\vec q) = {{\langle 0 |  J_i^X | V(\vec q)  \rangle} \over
 {\langle  0 |  J_i^Y | V(\vec q) \rangle}}.
\label{RRHO}
\ee
Note that $\langle 0 |  J_i | V)\rangle = \hat \epsilon_i m_V^2/f_V$
gives the vector meson decay constant $f_V$.
Naively speaking, $R^{X}_{\pi}=Z_{V^X}$ and $R^{X,Y}=Z_{V^Y}/Z_{V^X}$.

Eqn. \ref{RPI} is measured for $0<t<T$ with $T=10$ on $8^3\times 24$
lattices (20 for the hypercube and 40 for the clover action), on a subset
of the bare quark mass values,
and Eqn. \ref{RRHO} is a byproduct of spectroscopy.
The source at $t=0$ is a  Gaussian shell source centered on the origin;
the sink at $T$ is a point sink projected onto zero momentum, and
momentum was injected at the current. I was able to get
a signal for $\vec q = (2\pi/8) (1,0,0)$ in addition to $\vec q=0$.
Fits of  $R^{X,Y}_{\pi,\vec q= 0}$ are done by a correlated fit to the
two three-point functions, fits of
 $R^{X,Y}_{\pi,\vec q \ne 0}$ are done by single-elimination jacknife
and fits of $R^{X,Y}_\rho$ by a correlated fit to the two two-point functions.

The renormalization factors are determined sequentially. I begin with
a forward matrix element $R^{X}_{\pi}$ of the conserved current:
current conservation demands that this ratio be equal to unity, or
$Z_{J^C}=1$. Confirming that measurement, I can use Eqn. \ref{FWD}
or \ref{RPI} (at $\vec q=0$) to determine $Z_V$'s for the other
currents. Self consistency for lattice perturbation theory demands
that the $Z$'s should be process-independent, but in general that
will not be true: the $O(a)$ corrections to Eqn. \ref{CVC} will
make the ratios process-dependent. The absence of process-dependence in
the ratios of $Z$'s is a measure of improvement.

By itself, the ratio $R^{X}_{\pi}$ is contaminated by wrap-around
in time. The wrap-around mostly affects the two point
function and it can be corrected for, by fitting the two-point
function to a hyperbolic cosine ($\simeq A(\exp(-\mu t)+\exp(-\mu(N_t-t)))$)
and the three-point function to a forward-going exponential
$\simeq A/Z\exp(-\mu T)$.

\begin{figure}[thb]
\epsfxsize=0.8 \hsize
\epsffile{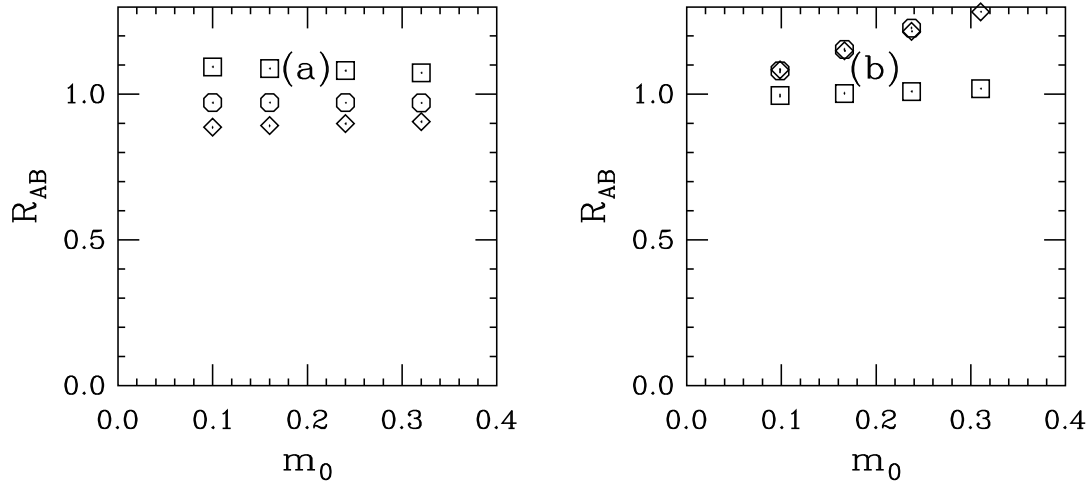}
\caption{
Ratios of three-point functions at $\vec p = 0$ from (a) the
hypercubic action and (b) the fat link clover action.
Octagons label $ R^{C,L}$, squares $R^{C,I}$, and diamonds
$R^{I,L}$. 
}
\label{fig:rats}
\end{figure}

\begin{figure}[thb]
\epsfxsize=0.8 \hsize
\epsffile{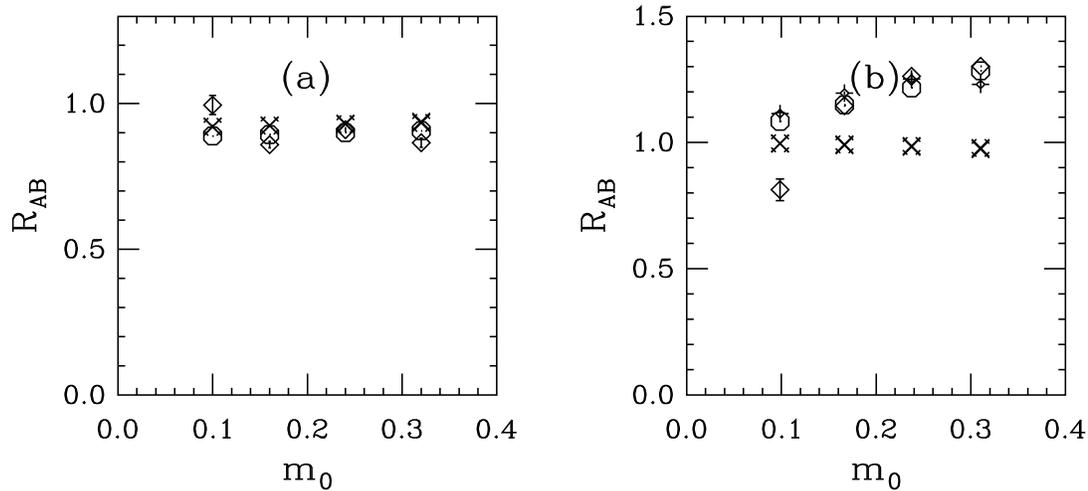}
\caption{$R^{I,L}$ from ratios of three point functions
(octagons $\vec q=0$, diamonds $\vec q=(2\pi/8,0,0)$,
fancy diamonds, $\vec q=(2\pi/8,2\pi/8,0)$) and two-point
functions (fancy crosses $\vec q=0$, crosses $\vec q=(2\pi/8,0,0)$).
(a) hypercubic action (b) fat link clover action.
}
\label{fig:ratsp}
\end{figure}

Tables \ref{tab:rpifp} and \ref{tab:rrhofp}
show the results for measurements of Eq. \ref{RRHO} and \ref{RPI} for
 the hypercube action, and the same results for the fat link clover
action are in Tables \ref{tab:rpicl} and \ref{tab:rrhocl}.
Figs. \ref{fig:rats} and \ref{fig:ratsp} illustrate the results.
 All the fat link $Z$-factors are close to unity.
In contrast, $Z_V^I \simeq 0.78$
 for the nonperturbatively-improved thin link  clover 
action at $\beta=6.0-6.2$ \cite{NPZ}.

Tables \ref{tab:rpifp} and \ref{tab:rpicl} show that the ratio of the
conserved current to either the local or improved currents is quite
different at $\vec q = 0$ and $\vec q \ne 0$. This is an artifact of
imperfection of the conserved current. Although the ratio of
the local current to the improved current is also $O(a)$, its
momentum dependence is much smaller. The conserved current also
does not respond well to being folded over into $\langle 0 | J_i|\rho\rangle$.
In Tables \ref{tab:rrhofp} and \ref{tab:rrhocl} I show only the
$\vec q=0$ ratio with the improved current: it is quite different from
the ratio of forward matrix elements.

Fig. \ref{fig:ratsp} shows a comparison of different measurements
of $Z_{V^L}/Z_{V^I}$ from two point and three point functions.
The scatter of the points for the hypercubic action is small and
appears to be mass-independent. In contrast, the Z-factors of
the fat link clover action
two-point and three-point functions are quite different, especially
at higher mass, although they appear to extrapolate into each other
at zero quark mass.  One could argue that since the local current
has $O(a)$ errors, and the improved current is $O(a^2)$, we are
just seeing an $O(a)$ effect and the improved current is the one to use.
However, the same argument should apply to the hypercubic action,
and there the effects are much smaller. If it is an $O(a)$ effect,
one will need more than a factor of two reduction in the lattice
spacing (or in $m_0 a$) to make the fat link clover action as
consistent as the hypercubic action.

\begin{table}
\begin{tabular}{|c|l|l|l|l|l|l|}
\hline
$am_q$
&  $R^{C,L}_{\pi,0}(\vec q = 0)$  & $R^{C,L}_{\pi,0}(\vec q = 2\pi/8)$   
&  $R^{C,I}_{\pi,0}(\vec q = 0)$  & $R^{C,I}_{\pi,0}(\vec q = 2\pi/8)$   
&  $R^{I,L}_{\pi,0}(\vec q = 0)$  & $R^{I,L}_{\pi,0}(\vec q = 2\pi/8)$ \\
\hline
   0.32  &0.970(1)  &1.01(2) & 1.074(1)& 1.19(1)   & 0.906(2) & 0.87(2) \\
   0.24  & 0.971(1) &1.11(2) & 1.081(1)& 1.24(1)  &0.899(2) &  0.91(1)  \\
   0.16  & 0.971(1) &1.03(3) & 1.088(1)& 1.24(1)  & 0.892(3) & 0.86(1) \\
   0.10  &0.971(1)  & 1.42(8) & 1.094(1)& 1.35(2)  &0.868(3) & 0.99(3)  \\
\hline
\end{tabular}
\caption{Table of $R^{X,Y}_{\pi,0}$ ,
 Gaussian hypercube action, $\beta=3.70$ ($aT_c=1/4$).}
\label{tab:rpifp}
\end{table}

\begin{table}
\begin{tabular}{|c|l|l|l|}
\hline
$\kappa$
&  $R^{C,I}_{\rho}(\vec q = 0)$ 
&  $R^{I,L}_{\rho}(\vec q = 0)$  & $R^{I,L}_{\rho}(\vec q = 2\pi/8)$ \\
\hline
   0.116 &0.574(9) & 0.933(1) & 0.938(2)  \\
   0.118 &0.621(13) & 0.930(1) & 0.931(2) \\
   0.120 &0.669(4) & 0.925(1) & 0.925(3) \\
   0.122 &0.732(7) & 0.920(1) & 0.921(4) \\
\hline
\end{tabular}
\caption{Table of $R^{X,Y}_\rho$ ,
 Gaussian hypercube action, $\beta=3.70$ ($aT_c=1/4$).}
\label{tab:rrhofp}
\end{table}

\begin{table}
\begin{tabular}{|c|l|l|l|l|l|l|}
\hline
$\kappa$
&  $R^{C,L}_{\pi,0}(\vec q = 0)$  & $R^{C,L}_{\pi,0}(\vec q = 2\pi/8)$   
&  $R^{C,I}_{\pi,0}(\vec q = 0)$  & $R^{C,I}_{\pi,0}(\vec q = 2\pi/8)$   
&  $R^{I,L}_{\pi,0}(\vec q = 0)$  & $R^{I,L}_{\pi,0}(\vec q = 2\pi/8)$ \\
\hline
   0.116 & 1.308(17) &1.349(4) & 1.136(2) &1.026(4) & 1.2383(1) & 1.301(1)  \\
   0.118 &1.227(3)  &1.449(3)) & 1.009(2) &1.254(4) & 1.216(1) &1.26(1) \\
   0.120 &1.116(3)& 1.262(12) &1.003(3) &  1.43(5) &1.149(2) & 1.15(1) \\
   0.122 &1.080(6) &  & 0.995(5)& & 1.084(1) &  \\
\hline
\end{tabular}
\caption{Table of $R^{X,Y}_{\pi,0}$ ,
 fat link clover action, $\beta=3.70$ ($aT_c=1/4$).}
\label{tab:rpicl}
\end{table}

\begin{table}
\begin{tabular}{|c|l|l|l|}
\hline
$\kappa$
&  $R^{C,I}_{\rho}(\vec q = 0)$    
&  $R^{I,L}_{\rho}(\vec q = 0)$  & $R^{I,L}_{\rho}(\vec q = 2\pi/8)$ \\
\hline
0.116 & 0.732(22) & 0.978(1) & 0.973(1)  \\
0.118 & 0.725(35) & 0.986(1) & 0.982(1) \\
0.120 & 0.714(18) & 0.993(1) & 0.989(1) \\
0.122 & 0.722(26) & 0.997(1) & 0.994(1) \\
\hline
\end{tabular}
\caption{Table of $R^{X,Y}_{\rho,0}$ ,
 fat link clover action, $\beta=3.70$ ($aT_c=1/4$).}
\label{tab:rrhocl}
\end{table}

\begin{figure}[thb]
\epsfxsize=0.8 \hsize
\epsffile{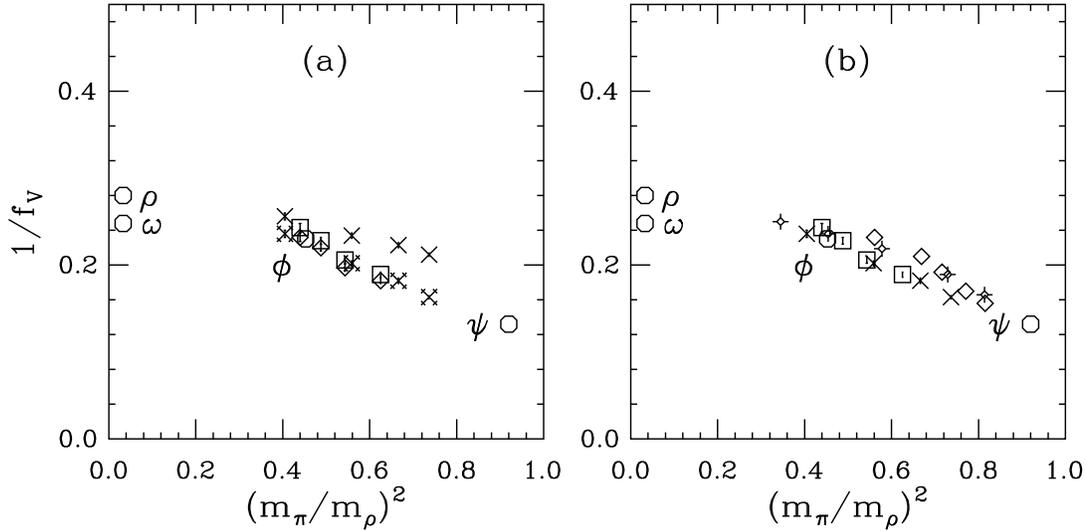}
\caption{(a)
Vector meson decay constants at $a=0.18$ fm from the
 hypercubic and fat link clover
actions. Diamonds and squares are the local and improved operator
with the hypercubic action while the crosses and fancy crosses are
the same operators with the fat link clover action.
(b) A comparison of $1/f_V$ using improved operators with
the hypercubic action (crosses), the fat link clover action (squares),
and the nonperturbatively tuned thin link clover action at
$\beta=6.0$ (diamonds) and 6.2 (fancy diamonds) from Ref. 26.}
\label{fig:fv}
\end{figure}

Finally, I can compare $1/f_V$ to the observed vector meson decay widths,
by multiplying the lattice value (given in Table \ref{tab:fvfp}
or \ref{tab:fvcl}) by the appropriate $Z$ factor
(extracted from the $\vec q=0$ $R^{C,X}_{\pi,0}$ correlator--
the first or third column of
 Table \ref{tab:rpifp} or \ref{tab:rpicl}). The result is
shown in Fig. \ref{fig:fv}a. All the currents bracket the $\phi$ meson
decay constant quite nicely, although there is considerable uncertainty in the
fat link clover result due to the different renormalization factors.
Vector decay constants have also been computed using the thin link clover
action, with a nonperturbatively tuned clover term,
at smaller lattice spacings (Wilson gauge action, $\beta=6.0$ and 6.2)
 in Ref. \cite{ALPHA}.
I plot a comparison between those values and the results of this simulation
(showing only the improved operator, to avoid clutter)
in Fig. \ref{fig:fv}b. There does not seem to be much to be gained for this
matrix element
by going to the smaller lattice spacings.

\begin{table}
\begin{tabular}{|c|l|l|}
\hline
mass &local & improved \\
\hline
0.32 & 0.188(2) & 0.177(2) \\
0.24 & 0.203(3) & 0.191(3) \\
0.16 & 0.226(3) & 0.210(3) \\
0.10 & 0.239(4) & 0.222(3) \\
\hline
\end{tabular}
\caption{Table of lattice $1/f_V$ for the hypercubic action.
}
\label{tab:fvfp}
\end{table}

\begin{table}
\begin{tabular}{|c|l|l|}
\hline
$\kappa$ &local & improved \\
\hline
0.116 &  0.162(2) & 0.159(2) \\
0.118 & 0.181(2) & 0.179(2) \\
0.120 & 0.203(4) & 0.201(3) \\
0.122 & 0.236(4) & 0.235(4) \\
\hline
\end{tabular}
\caption{Table of lattice $1/f_V$ for the fat link clover action.
}
\label{tab:fvcl}
\end{table}

\subsection{The Axial Current}
I have performed a nonperturbative determination of the axial vector
current renormalization factor following the classical approach of
Martinelli and Maiani\cite{MM}. This begins with the Ward identity
Eqn. \ref{WARD1}, with the operator $O = A_\nu^b(y) V_\rho^c(z)$
where
$V_\rho^c(z)= \bar \psi(z) {\lambda^c \over 2} \gamma_\rho \psi(z)$ and
$A_\nu^b(y)= \bar \psi(z) {\lambda^b \over 2} \gamma_\nu \gamma_5 \psi(y)$.
Standard manipulations lead to the identity
\bee
2m_q \sum_n \sum_{\vec y} \langle P^a(n)A_\nu^b(\vec y, y_0) V_\rho^c(0,0) =
-if^{abd} \sum_{\vec y} \langle V_\nu^d(\vec y,y_0)V_\rho^c(0,0) \rangle
-if^{acd}\sum_{\vec y}\langle A_\nu^b(\vec y,y_0)A_\rho^d(0,0) \rangle.
\ee
($P^a(n)$ is the pseudoscalar current 
$\bar \psi(n) {\lambda^a \over 2} \gamma_5 \psi(n)$.)
To pass to a lattice description, all currents are rescaled $J \rightarrow
Z_L J_L$ and the quark mass is replaced by the unnormalized quark mass
from the lattice PCAC relation,
$m_q \rightarrow \rho =m_q Z_P/Z_A $ with
\bee
 2\rho = {{ \sum_{\vec y}\langle \partial_0 A_0(\vec y, y_0) C(0,0) \rangle}
\over { \sum_{\vec y}\langle P(\vec y, y_0) C(0,0) \rangle}}
\label{RHO}
\ee
computed through a (correlated) fit to two two-point functions. The
lattice Ward identity then becomes
\bee
-2 \rho \langle PAV\rangle = {Z_V \over Z_A }\langle VV \rangle -
 {1\over Z_V }\langle AA \rangle.
\ee
(I computed $Z_A$ for the local axial current; the Ward identity for
the improved current involves additional contact terms.)
As for the vector current, I evaluated the two point functions
and the three point function with a source on $t=0$ and a sink on $t=y_0=10$.
I used 40 hypercubic propagators and 80 fat link clover propagators,
taking the local $Z_V$ from the forward current matrix element (rightmost data
column of Tables \ref{tab:rpifp} and \ref{tab:rpicl}).
I only looked at three lighter quark masses.
Correlated fits to the appropriate two point functions, Eqns. \ref{RHO}
and a correlated fit extracting
 $\langle 0 | A_0 | PS \rangle = f_{PS} m_{PS}$, produced
 the results shown in Tables \ref{tab:fpifp} and \ref{tab:fpicl}. The
 two errors on $Z_A$
are from the jacknife and from the variation of $\rho$ and $Z_V$.
I use $a \sqrt{\sigma}=0.4049(37)$ from Ref. \cite{HYP1} to connect
the lattice spacing $a$ and string tension $\sigma$.
In the last column, I quote a physical number using a nominal
 $\sqrt{\sigma}=440$ MeV. A linear extrapolation to the chiral limit
in the quark mass produces the bottom line. The agreement with
experiment from the hypercube action for the pion seems acceptable.
Notice that $Z_A$ shows mild mass dependence and that, in the chiral limit,
it is also quite close to unity. The clover action shows smaller
mass dependence. In contrast, the nonperturbatively-tuned thin link clover
action $Z_A$ from the rotated operator is 0.79 at Wilson gauge action
 $\beta=6.0$ \cite{NPZ}

\begin{table}
\begin{tabular}{|c|l|l|l|l|}
\hline
mass& $af_{PS}$ & $Z_A^L$ & $f_{ps}/\sqrt{\sigma}$ & $f_{ps}$, MeV \\
\hline
0.24 & 0.170(2) & 0.865(1)(1) & 0.3632(55) & 160(3)  \\
0.16 & 0.156(2) & 0.902(2)(2) & 0.3475(60)  & 153(3) \\
0.10 & 0.147(2) & 0.916(2)(2) & 0.3326(55)    &    146(3)    \\
\hline
0    &          & 0.966(3)        &             &            134(6)\\
\hline
\end{tabular}
\caption{Table of lattice axial parameters for the hypercube action.
}
\label{tab:fpifp}
\end{table}

\begin{table}
\begin{tabular}{|c|l|l|l|l|}
\hline
$\kappa$& $af_{PS}$ & $Z_A^L$ & $f_{ps}/\sqrt{\sigma}$ & $f_{ps}$, MeV \\
\hline
0.118 & 0.135(2) & 0.851(33)(4) & 0.283(13) & 124(7)  \\
0.120 & 0.134(2) & 0.870(79)(4) & 0.288(30)  & 127(13) \\
0.122 & 0.130(3) & 0.872(28)(4) & 0.280(12) & 123(5)  \\
\hline
0    &           & 0.89(5)        &             &  122(10)\\
\hline
\end{tabular}
\caption{Table of lattice axial parameters for the fat link clover  action.
}
\label{tab:fpicl}
\end{table}

\section{Conclusions}
Few of the results shown here are surprising. The hypercube action
 is designed
to show improved chiral and kinetic properties, and it does.
The fat link clover action has poorer kinetic properties, but
in the small mass limit
they can be compensated for by going to smaller lattice spacing.
Both actions have vector and axial current renormalization factors which
are very close to unity.

It is  an open question to me, whether the hypercube action
is a practical improvement over the simpler alternative of
 the fat link clover action.
The cost of the action is a factor of 17 per computation
of $\psi=\Delta \chi$ compared to the usual clover action, with a gain
that typically only half as many steps of the inversion algorithm
 are required to construct $\Delta^{-1}$. The fat link clover action
\cite{FATCLOVER} has all the gain with no additional cost. Of course,
its kinetic properties are not improved, and so one must reduce the
lattice spacing to improve the dispersion relation and scaling
behavior of matrix elements. Heavy quark physics with the fat link clover
action will suffer from the same kinetic artifacts as the standard
clover action. 

A potential use for the hypercube
 action  is to construct an action which realizes the
Ginsparg-Wilson exactly. That is an active area of research \cite{EXACTGW},
with most attempts involving an iteration using the Wilson action as
a zeroth-order action. The eigenmodes of a GW action lie on or
 close to a circle.
The thin link Wilson action, or even the clover action, seem to be
poor choices for a zeroth-order action, because their free-field
eigenvalue spectrum does not look anything like a circle, and because
their real eigenmode spectrum in background instanton configurations
does not look anything like a step function (recall Fig. \ref{fig:mvsrho}).
The fat link stiffens the real eigenmode spectrum on instantons.
A hypercube action like the one described here looks like a much
 better choice, provided of course
the cost of its use is less than the gain in number of iteration steps.

Finally, dynamical fermion simulations with fat link actions still
appear \cite{DYNAMICAL} to require smaller levels of fattening than
are needed to substantially improve the chiral properties
of Wilson-type actions. 
 But for quenched calculations either action seems to me to be
superior to a thin link action.

\section*{Acknowledgements}
I would like to thank
Anna Hasenfratz, Tamas Kovacs,
and the members of the MILC Colaboration
for useful conversations and C. Sachrajda and
 J. Skullerud for instructive correspondence.
This work was supported by the U.S. Department of 
Energy, with some computations done on the T3E at Pittsburgh Supercomputing
Center through resources awarded to the MILC collaboration,
and on the 
Origin 2000 at the University of California, Santa Barbara.

\section*{Appendix: Blocking out of the Continuum}
A particularly interesting choice of blocking kernel for blocking
out of the continuum is the
overlapping transformation
\bee
\Omega(q)=\exp(-{1\over 2}c^2 q^2).
\ee
In my work, blocking out of the continuum was approximated
by beginning with  free Dirac fermions
on a fine lattice, making one RG step with a blocking factor $F=8$,
 and blocking onto a coarse lattice of size $5^4$ sites.   

\begin{figure}[thb]
\epsfxsize=0.8 \hsize
\epsffile{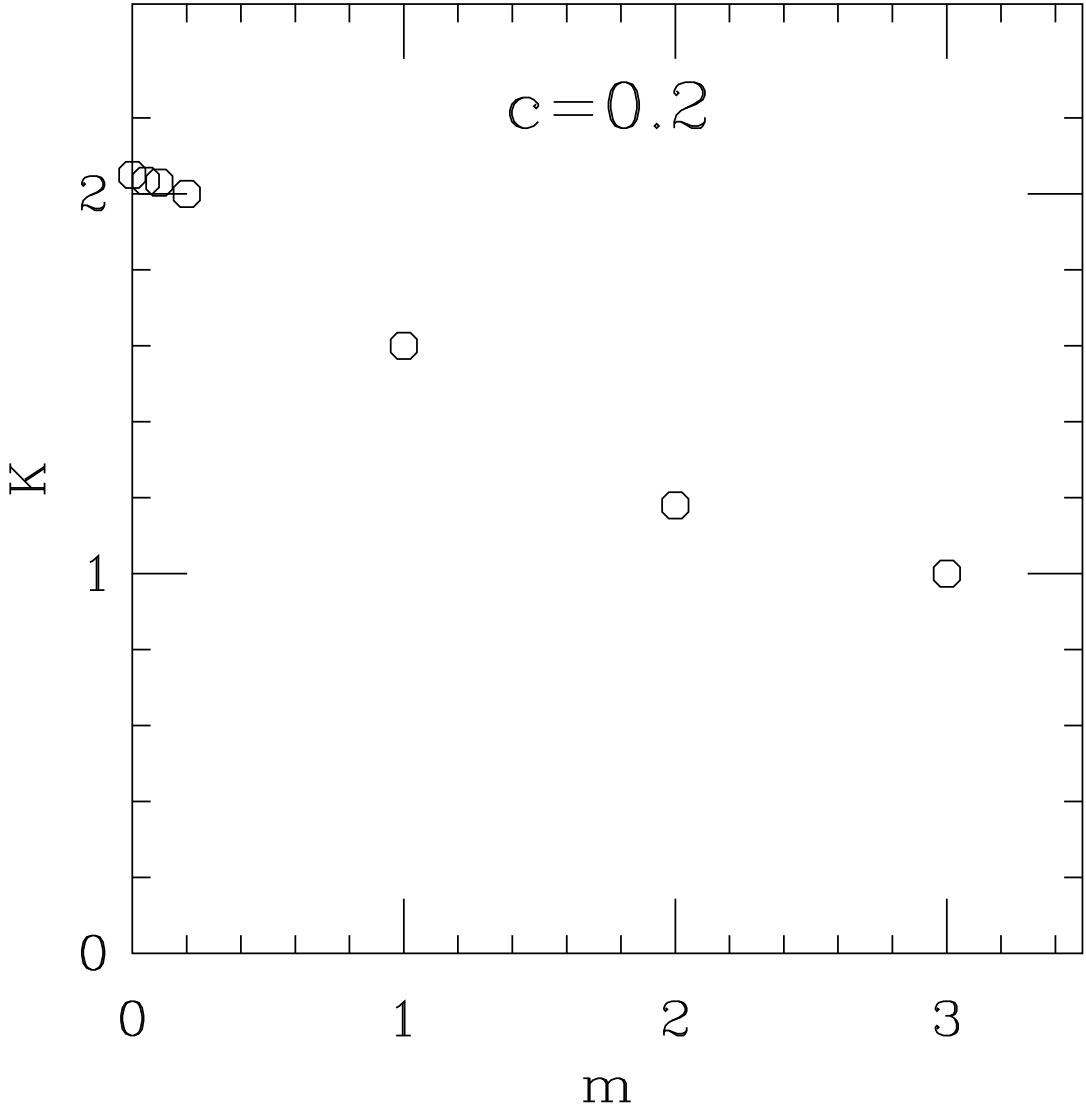}
\caption{
Optimum $K$ defining the RT of the massive action.}
\label{fig:avsm}
\end{figure}

Truncating the action to a hypercube and requiring good behavior
for the resulting free hypercubic action leads to the parameter
choice $c=0.2$ and $K=2.05$. By beginning with a bare mass $m/F$
on the fine scale I constructed a renormalized trajectory of
free massive fermions.  The parameterization and dispersion relation
was well behaved out to very large mass ($am \simeq 3.0$). The optimum
$K$ fell smoothly with increasing $m$, to a value of 1.0 at $m=3$.
This is shown in Fig. \ref{fig:avsm}.
In the actual simulations, the hypercubic parameters were fit
to the forms $\rho_j(m)$, $\lambda_j(m) = a\exp(- b m - c m^2)$
and the on-site coupling was fixed using the functional relation between
the $\rho$'s, $\lambda$'s, and bare mass $m$. If negative bare masses
are needed, I extrapolated
exactly as described in Ref. \cite{HYP1}.The action is listed in Table
\ref{tab:couplings}. This is of course not a unique (or probably even an
optimal) parameterization.

\begin{table}
\begin{tabular}{|c|l|}
\hline
  & $\lambda(r)$ \\
\hline
offset & \\
\hline
1 0 0 0 &    $ -0.0725 \exp(-0.7092m - 0.0149m^2)$  \\
1 1 0 0 &    $  -0.0319 \exp(-1.007m -0.0421m^2)$  \\
1 1 1 0 &    $  -0.0156 \exp(-1.123m -0.0803m^2)$ \\
1 1 1 1 &    $   -0.0080\exp(-1.180m-0.1169m^2)$ \\
\hline
  & $\rho_0(r)$ \\
\hline
offset & \\
\hline
1 0 0 0 &    $  -0.1819 \exp(-0.9475m + 0.0031m^2)$ \\
1 1 0 0 &    $  -0.0318 \exp(- 1.031m-0.0502m^2)$ \\
1 1 1 0 &    $  -0.00897 \exp(-1.017m-0.1047m^2)$  \\
1 1 1 1 &    $   -0.00295\exp(-0.9307m-0.1633m^2)$ \\
\hline
\end{tabular}
\caption{Couplings of the hypercubic action used in this
work. The free action is parameterized as
 $\Delta = \lambda(r)+ i \gamma_\mu \rho_\mu(r)$.
}
\label{tab:couplings}
\end{table}

\begin{table}
\begin{tabular}{|c|l|}
\hline
  & $\lambda(r)$ \\
\hline
offset & \\
0 0 0 0 & $0.1047 + 0.3287m-0.0639m^2$ \\
1 0 0 0 &    $  \exp(-3.4070-0.6776m- 0.0639m^2 )$ \\
1 1 0 0 &    $   \exp(-4.2303 -0.9708m -0.07609m^2 )$ \\
1 1 1 0 &    $   \exp (-4.9495-1.0860m -0.1147m^2)$ \\
1 1 1 1 &    $   \exp( -5.6120 -1.1437m-0.1509m^2)$ \\
\hline
  & $\rho_0(r)$ \\
\hline
offset & \\
\hline
1 0 0 0 &    $   \exp(-2.4746-0.9150m-0.09179m^2 )$ \\
1 1 0 0 &    $   \exp(-4.2383-0.9947m-0.0846m^2 )$ \\
1 1 1 0 &    $   \exp(-5.5061-0.9809m-0.1388m^2 )$\\
1 1 1 1 &    $   \exp(-6.6218-0.8943m-0.1969m^2)$ \\
\hline
\end{tabular}
\caption{The hypercubic approximation to the FP field
for the hypercubic action used in this
work. The field is parameterized as
 $\zeta = \lambda(r)+ i \gamma_\mu \rho_\mu(r)$.
}
\label{tab:fpop}
\end{table}

The free FP field was constructed by summing Eqn. \ref{FPFIELD}, in parallel
with the FP action. Again, I truncated the FP field to a hypercube
and parameterized the ``offset'' terms exactly as for the hypercubic
action. These couplings are shown in Table \ref{tab:fpop}.
Both the action and FP field are made gauge invariant by averaging
the gauge connections over all the shortest paths,
 exactly as in Ref. \cite{HYP1}

\begin{figure}[thb]
\epsfxsize=0.8 \hsize
\epsffile{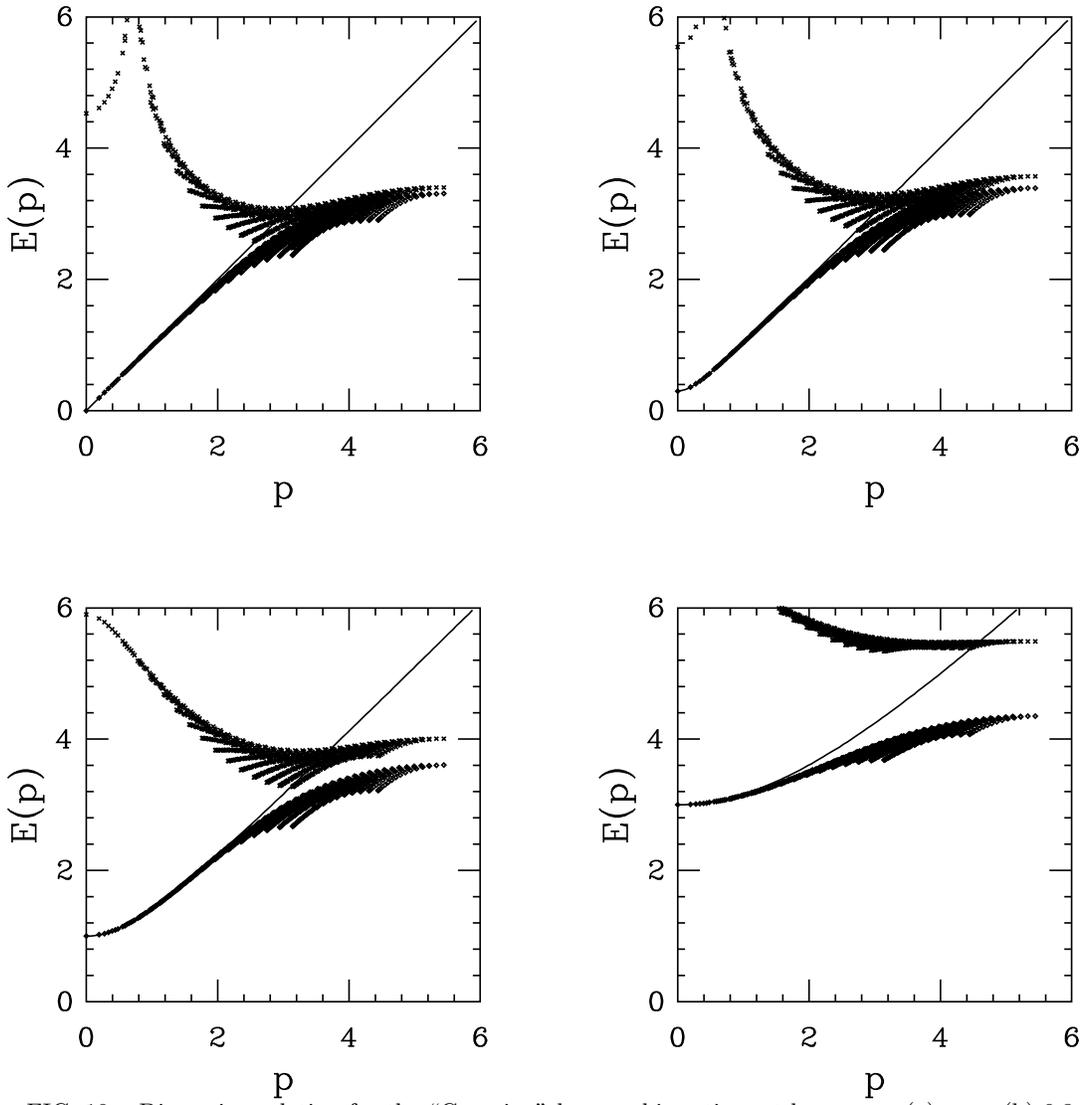}
\caption{
Dispersion relation for the ``Gaussian'' hypercubic action,
at bare mass (a) zero, (b) 0.3, (c) 1.0, (d) 3.0. The line is
the continuum expectation.}
\label{fig:alldr}
\end{figure}

\begin{figure}[thb]
\epsfxsize=0.8 \hsize
\epsffile{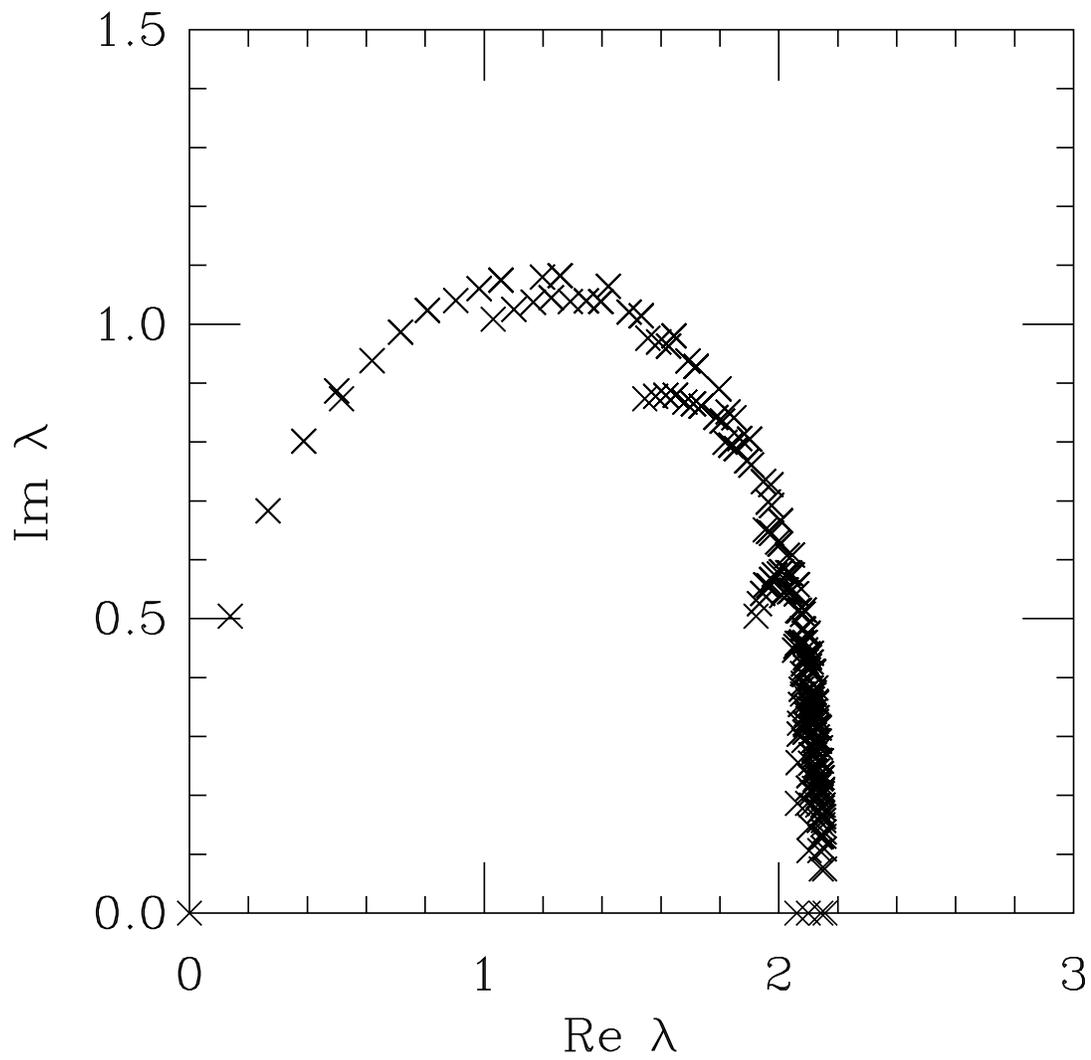}
\caption{
Eigenmode spectrum of the ``Gaussian'' hypercubic action.}
\label{fig:eigeng}
\end{figure}

\begin{figure}[thb]
\epsfxsize=0.8 \hsize
\epsffile{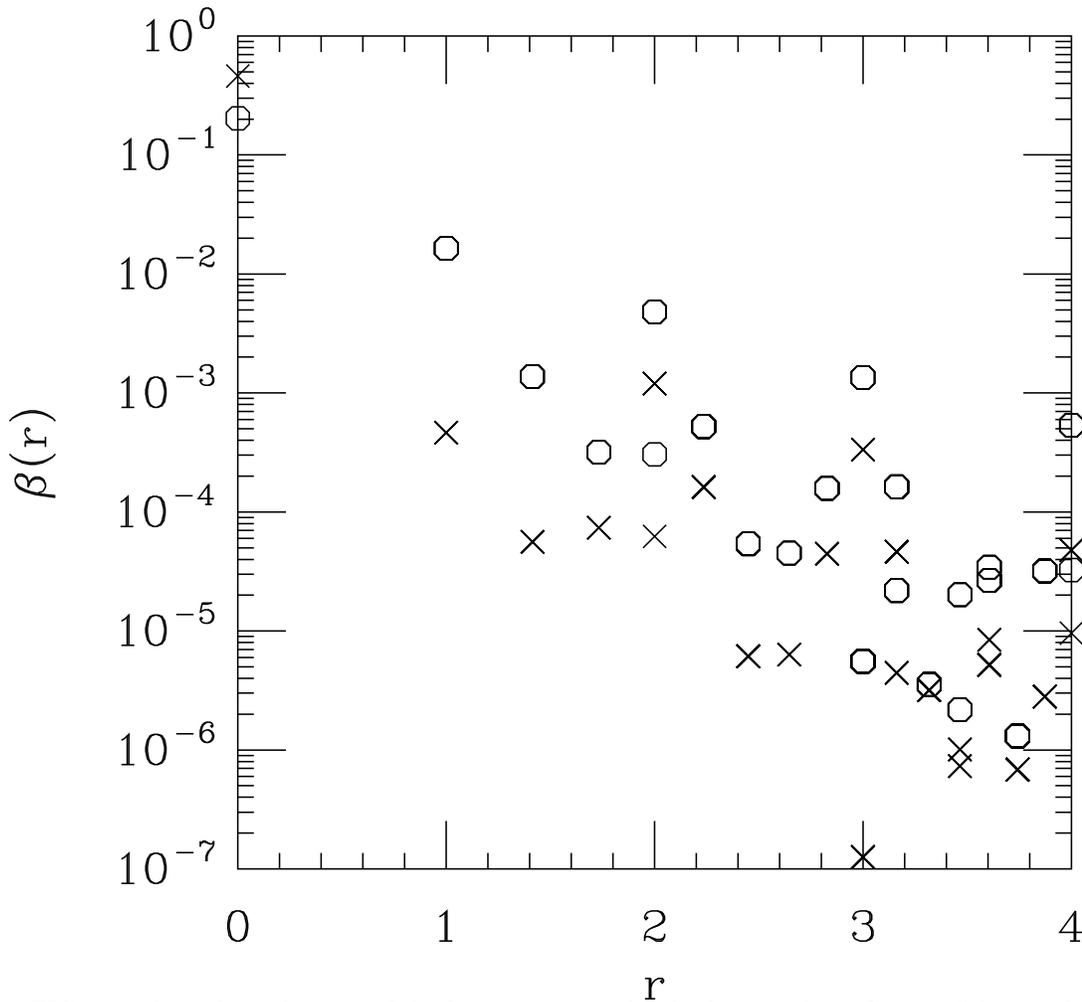}
\caption{
$\beta(r)$, the scalar part of the free propagator, for
the free massless ``Gaussian'' hypercubic action (crosses) and for the Wilson
action (octagons).}
\label{fig:beta0001}
\end{figure}

I show a few properties of the free field action:
Fig. \ref{fig:alldr} shows dispersion relations of the hypercubic
action on a large lattice, for several values of the bare mass.
In Fig. \ref{fig:eigeng} I show the eigenmode spectrum of the
hypercubic Gaussian action (with only
the positive imaginary part of the complex eigenmodes). The eigenmodes
of the FP action  lie on a circle of radius $K/2$ centered
at the point $(K/2,0)$, and the approximate action tracks the
circle closely.
Finally, in  Fig. \ref{fig:beta0001} I display the scalar part of
the free fermion propagator,
 $\Delta(r)^{-1} = \beta(r) + i \gamma_\mu \alpha_\mu(r)$. The 
Ginsparg-Wilson relation for the propagator probes 
$1/2\{\gamma_5,\Delta(r)^{-1}\}= \gamma_5 \beta(r)$.
This is shown for massless fermions on an $11^4$ lattice with
antiperiodic boundary conditions. For the real FP action,
$\beta(r)$ should be proportional to a delta-function at the origin.
For the hypercubic action it lies about 1-2  orders of magnitude below
the value of the Wilson action at all nonzero $r$.

\end{document}